\input amstex
\documentstyle{amsppt}
\Monograph
\font\nf=wncyr10

\font\rit=wncyi10
\nf

\def\sq {\square}
\def\dim{\operatorname {dim}}
\def\Im {\operatorname {Im}}
\def\JU{\char'020}\def\ji{\char'032}\def\JA{\char'027}\def\ja{\char'037}\def\m{\char'176}
\NoBlackBoxes

\baselineskip 12pt

\bigskip

\bigskip

\centerline {\bf THE WAGNER CURVATURE TENSOR  IN NONHOLONOMIC MECHANICS}

\bigskip

{\rm
\centerline {Vladimir Dragovi\' c\footnote{vladad\@mi.sanu.ac.yu, Mathematical Institute, Belgrade, Yugoslavia \& SISSA, Trieste, Italy} and Borislav
Gaji\' c\footnote{gajab\@mi.sanu.ac.yu, Mathematical Institute, Belgrade, Yugoslavia}}

\

\bigskip
\centerline {\bf Abstract}
\bigskip

We present the classical Wagner construction from 1935 of the
curvature tensor for completely nonholonomic manifolds in both
invariant and coordinate way. The starting point is the Shouten
curvature tensor for nonholonomic connection introduced by
Vranceanu and Shouten. We illustrate the construction on two
mechanical examples: the case of a homogeneous disc rolling
without sliding on a horizontal plane and the case of a
homogeneous ball rolling  without sliding on a fixed sphere. In
the second case we study the conditions  on the ratio of diameters
of the ball and the sphere to obtain a flat space - with the
Wagner curvature tensor equal zero.

\bigskip
\bigskip

\centerline {\bf Table of contents}

\

\ \ \ \S 1. Introduction.

\ \ \ \S 1.1. A historical overview.

\ \ \ \S 1.2. Basic notions from nonholonomic geometry.

\ \ \ \S 1.3. The equations of motion of  mechanical nonholonomic systems.

\ \ \ \S 2. The Shouten tensor.

\ \ \ \S 3. The Wagner tensor.

\ \ \ \S 3.1. The Wagner construction.

\ \ \ \S 3.2. Coordinate expression for the Wagner tensor.

\ \ \ \S 3.3. Absolute parallelism and the Wagner tensor.

\ \ \ \S 4. The disc rolling on the plane.

\ \ \ \S 5. The ball rolling on the sphere.

\ \ \ \ Conclusion.

\ \ \ \ Acknowledgement.

\ \ \ \ References.

\newpage

\

\centerline {\bf \S 1. Introduction}

\

\centerline {\bf 1.1. Historical overview}

\

It is well known that the full difference between nonholonomic
variational problems and nonholonomic mechanics was understood
after Hertz [Hr]. The geometrization of nonholonomic mechanics
started in late 20' of the XX century, with works of Vranceanu,
Synge and Shouten. Vranceanu  defined the notion of nonholonomic
structure on a manifold (see [Vr]). Synge and Shouten made the
first steps toward the definition of the curvature in nonholonomic
case (see [Sy, Sh]). It was Shouten who introduced the notion of
partial, or nonholonomic connection. However, the highlights of
that pioneers period of development of mechanically motivated
nonholonomic geometry was the work of V. V. Wagner, published in
several papers from 1935 till 1941 (see [Wa1, Wa2, Wa3]). Wagner
constructed the curvature tensor as an extension of the Shouten
tensor. This construction is performed in several steps, following
the flag of the distribution. In that sence, the structure of
nonholonomicity of given distribution is reflected in the Wagner
construction. For those achievements, Wagner was awarded by Kazan
University in 1937 (see [VG]).

The main aim of this paper is to present Wagner's construction,
both in invariant and coordinate way. The existence of
Gorbatenko's recent, modern review [Go] is very helpful in
understanding original Wagner's works. Since we want to follow the
original Wagner ideas, there are some differences from
Gorbatenko's presentation.

We also give two mechanical examples. The first one is  the
problem of a homogeneous disc rolling  without sliding on a
horizontal plane and the second is the problem of a homogeneous
ball rolling  without sliding on a fixed sphere. In both cases we
produced complete computations of the construction of the Wagner
curvature tensor. Although the first problem is of degree 2 of
nonholonomicity, and the second one is of degree 1, the
computations in the second case are much more complicated.

The problem of homogeneous ball rolling  without sliding on a
fixed sphere is interesting because it gives a family of $(3, 5)$-
problems depending on a parameter $k$, which is the ratio between
the diameters of the ball and the sphere. We investigate the
Wagner flatness in these cases, in terms of this parameter $k$.

Geometry of nonholonomic variational problems is intensively
developing nowadays, (see  [Ju, Mn, AS]) motivated by the Control
Theory. As an important example, we mention the Agrachev curvature
tensor and related invariants of Sub-Riemannian Geometry (see
[AS]). These natural geometric constructions were developed
further in [AZ1, AZ2], and Agrachev and Zelenko implied their
theory to the situation of a homogeneous ball rolling  without
sliding on a fixed sphere. It appears that there exist $k$ for
which their invariants are zero, exactly in the same cases where
the Cartan tensor is zero (see [Ca,
 Mn]).

 So, putting alltogether, we can summarize the conclusion of this paper
by saying that the Wagner construction of curvature tensor is
natural, and essentially different from  other natural
constructions, such as the Cartan and the Agrachev curvatures.

\

\

\centerline {\bf 1.2. Basic notions from nonholonomic geometry}

\bigskip

{\rm
Let us fix some basic notions from the theory of distributions [VG]..
\vskip 0.2cm
{\bf Definition 1.} Let $\displaystyle TM=\bigcup_{x \in M} T_xM$, be the
tangent
bundle of a smooth  $n$-dimensional manifold $M$. A sub-bundle
$\displaystyle V=\bigcup_{x\in M}V_x$, where  $V_x$ is a vector subspace of
$T_xM$, smoothly dependent on points  $x \in M$, is a  {\it distribution}.
If the manifold $M$ is connected $\dim V_x$ is called {\it the dimension} of the
distribution.
\vskip 0.2cm

\vskip 0.2cm

 A vector field $X$ on $M$ {\it belongs} to the distribution $V$ if $X(x)
\subset V_x$. A curve  $\gamma$ is {\it admissible relatively to } $V$,
if the  vector field $\dot \gamma$ belongs to $V$.
\vskip 0.2cm

{\it  A differential system} is a  linear space of vector fields
having a structure of $C^{\infty}(M)$ - module. Vector fields which belong
to the distribution $V$ form a differential system $N(V)$.
\vskip 0.2cm

A $k-$ dimensional distribution $V$ is {\it integrable} if the manifold
$M$ is foliated to $k-$ dimensional sub-manifolds, having $V_x$ as the
tangent space at the point $x$. According to the Frobenius theorem,
$V$ is integrable if and only if the corresponding differential
system $N(V)$ is {\it involutive}, i.e. if it is a Lie sub-algebra
of Lie algebra  of vector fields on $M$.
\medskip

{\bf Definition 2.} {\it The  flag} of a differential system $N$ is a sequence of
differential systems: $N_0=N,\ N_1=[N,N],
\dots,\ N_l=[N_{l-1},N], \dots $.
\vskip 0.2cm

The differential systems $N_i$ are not always differential systems of some
 distributions $V_i$,
but if for every  $i$, there exists $V_i$, such that $N_i=N(V_i)$, then there
exists
{\it a flag} of the distribution $V$: $V=V_0\subset V_1\dots$. Such distributions, which have  flags, will be called {\it  regular}. It is clear that the sequence  $N(V_i)$
is going to stabilize, and there exists a number $r$ such that $N(V_{r-1})\subset N(V_r)=N(V_{r+1})$.
\vskip 0.2cm
{\bf Definition 3.} If there exists a number $r$ such that $V_r=TM$, the distribution $V$ is called
{\it completely  nonholonomic}, and minimal such $r$ is {\it the degree of nonholonomicity} of the  distribution $V$.
\vskip 0.2cm
We are going to consider only
regular and completely  nonholonomic distributions.

\

\

\centerline {\bf 1.3. The equations of motion of  mechanical nonholonomic systems}

\

One of the basic references on nonholonomic mechanics is [NF], see also [AKN].
Let us consider nonholonomic mechanical system corresponding to a Riemannian
manifold $(M,g)$, where $g$
is a metric defined by the  kinetic energy. It is well-known that to every Riemannian
metric $g$ on $M$ corresponds a connection $\nabla$  with the properties :
$$
\aligned
{\text\rm i)}&\quad \nabla _Xg(Y,Z)=X(g(Y,Z))- g(\nabla_XY,Z)-g(Y,\nabla_XZ)=0,\\
{\text\rm ii)}&\quad T(X,Y)=\nabla_XY-\nabla_YX-[X,Y]=0,
\endaligned
$$
where $X, Y, Z$ are smooth vector fields on  $M$.
This {\it symmetric, metric} connection is usually called  {\it the  Levi-Chivita} connection.

We assume that the distribution $V$  is  defined
by  $(n-m)$ 1-forms $\omega_\alpha$;
in  local coordinates $q=(q^1,...,q^n)$ on $M$
$$
\omega_\rho (q)(\dot q)=a_{\rho i}(q) \dot q^i=0 \quad
\rho=m+1,\dots ,n\quad ;\quad i=1,\dots ,n.
\tag1
$$
{\bf Definition 4.} {\it A virtual displacement} is a vector field $X$ on $M$, such that  $\omega_\rho(X)=0$, i.e.  $X$ belongs to the  differential  system $N(V)$.
\vskip 0.2cm
Differential equations of motion of a given mechanical system follow from the  D'Alambert-Lagrange
principle: {\it trajectory $\gamma$ of given system is a solution of the equation
$$
\langle \nabla_{\dot \gamma}\dot {\gamma}-Q,X\rangle =0,
\tag2
$$
where  $X$ is an arbitrary virtual displacement, $Q$ a vector field of internal forces, and $\nabla$ is the metric connection for the metric $g$.}

The vector field $R(x)$ on $M$, such that
$R(x)\in V_x^\perp ,\ V_x^\perp\oplus V_x=T_xM$, is called
{\it reaction of  ideal nonholonomic connections}.
Equation (2) can be written in the form:
$$
\aligned
&\nabla_{\dot \gamma}\dot {\gamma}-Q=R,\\
&\omega_\alpha(\dot\gamma)=0.
\endaligned
\tag3
$$
If the system is potential, by introducing $L=T-U$, where  $U$ is the potential
 energy of the  system ($Q=-grad U$), then in local coordinates
$q$ on $M$, equations (3) become:
$$
\aligned
\frac d{dt}\frac {\partial L}{\partial \dot q}-\frac {\partial L}{\partial q}=
\tilde{R},\\
\omega_\alpha(\dot q)=0.
\endaligned
\tag4
$$
Now  $\tilde{R}$ is a 1-form in $(V^\perp)$, and it can be represented as
a linear combination of  1-forms $\omega^{m+1},\dots, \omega^n$ which define
the distribution: $\tilde{R}=\sum\limits_{\alpha=m+1}^n \lambda_\alpha \omega_\alpha$.

Suppose   $e_1,\dots,e_n$ are the vector fields on  $M$, such that $e_1(x),\dots ,
e_n(x)$ form a  base of the  vector space $T_xM$ at every point $x\in M$, and
$e_1,\dots,e_m$ generate the  differential system $N(V)$.
Express them through the  coordinate vector fields:
$$
e_i=A_i^j(q)\frac{\partial}{\partial q^j}, \quad i,j=1,\dots,n.
$$

Denote by $p$ a projection $p:TM \rightarrow V$ orthogonal
according to the  metric $g$. Corresponding homomorphism of
$C^\infty$-modules of sections of $TM$ and  $V$ will be also
denoted by $p$:
$$
p\left( \frac{\partial}{\partial q^i} \right) =p_i^ae_a, \quad a=1,\dots,m,
\quad i=1,\dots ,n .
$$
Projecting by  $p$ the equations  (3), from $R(x)\in V^\perp (x)$, we get
$p(R)=0$, and denote  $p(Q)=\tilde Q$ we get
$$
\tilde {\nabla}_{\dot\gamma}\dot \gamma=\tilde Q,
\tag5
$$
where  $\tilde \nabla$ is the projected connection. A relationship between  coefficients  $\tilde {\Gamma}^c_{a b}$ of the connection $\tilde \nabla$,
defined by the formula
$$
\tilde {\nabla}_{e_a}e_b=\tilde {\Gamma}^c_{a b}e_c
$$
and  the Christoffel symbols $\Gamma^k_{i j}$ of the connection $\nabla$ follows from
$$
\aligned
\tilde{\nabla}_{e_a}e_b=\tilde {\Gamma}^c_{a b}e_c&=p\left(\nabla_{e_a}e_b\right)\\
&=p\left(\nabla_{A_a^i\frac {\partial}{\partial q^i}}A^j_b
\frac {\partial}{\partial q^j}\right)\\
&=p\left( A_a^i\frac {\partial A^j_b}{\partial q^i}\frac {\partial}
{\partial q^j}+A_a^iA_b^j\nabla_{\frac {\partial}{\partial q^i}}
\frac {\partial}{\partial q^j}\right)\\
&=A_a^i \frac {\partial A_b^j}{\partial q^i}p_j^ce_c+A_a^iA_b^j
\Gamma^k_{i j}p^c_ke_c.
\endaligned
$$
Thus we get
$$
\tilde {\Gamma} ^c_{a b}=\Gamma^k_{i j}A_a^iA_b^jp_k^c+A_a^i
\frac {\partial A_b^j}{\partial q^i}p_j^c.
\tag{6}
$$

If the motion is taking place under the inertia ($Q=\tilde Q=0$), the  trajectories of
nonholonomic mechanical problem are going to be geodesics  for the
projected connection $\tilde\nabla$.
Equations (5) were derived by Vrancheanu and  Shouten.
\medskip

{\bf Note.} The projected connection $\tilde\nabla$ is not a connection on the
vector bundle $V$ over $M$, because  the  parallel
transport is defined only along  admissible curves.
So,  it is called {\it partial} or {\it  nonholonomic}
connection.
(Exact definition follows in Section 2.2).

\

\

\centerline {\bf \S 2. THE SHOUTEN TENSOR}

\

Let $V$ be a distribution on $M$. Denote  $C^\infty(M)$-
module of sections on $V$ by  $\Gamma (V)$.
\vskip 0.2cm
{\bf Definition 1.} {\it A nonholonomic connection } on the sub-bundle $V$ of $TM$ is a map $\nabla :\Gamma(V)\times \Gamma(V)\rightarrow \Gamma(V)$ with the properties:
$$
\aligned
{\text \rm i)}&\quad \nabla_X(Y+Z)=\nabla_XY+\nabla_XZ\\
{\text \rm ii)}&\quad \nabla_X(f\cdot Y)=X(f)Y+f\nabla_XY\\
{\text \rm iii)}&\quad \nabla_{fX+gY}Z=f\nabla_XZ+g\nabla_YZ\\
&X,Y,Z \in \Gamma(V)\quad;\quad f,g\in C^\infty(M).
\endaligned
$$
\vskip 0.2cm Having a morphism of  vector bundles $p_0:TM
\rightarrow V$, formed by the  projection on $V$, denote by
$q_0=1_{TM}-p_0$ the projection on $W$, $V\oplus W=TM$. \vskip
0.2cm

{\bf Definition 2.} The tensor field $T_\nabla:\Gamma(V)\times\Gamma(V)
\rightarrow\Gamma(V)$ defined in the following way:
$$
T_\nabla(X,Y)=\nabla_XY-\nabla_YX-p_0[X,Y]\quad ;\quad X,Y\in\Gamma(V)
$$
is called  {\it the tensor of torsion} for the connection $\nabla$.
\vskip 0.2cm
Suppose there is  a positively defined metric tensor $g$ on
$V$:
$$
g:\Gamma(V)\times\Gamma(V)\rightarrow C^\infty(M),\quad g(X,Y)=g(Y,X).
$$
\vskip 0.2cm

{\bf Theorem 1.} {\it Given  a distribution $V$, with $p_0$ and $g$, there exists a unique  nonholonomic connection $\nabla$ with the properties:
$$
\aligned
{\text\rm i)}&\quad \nabla_Xg(Y,Z)=X(g(Y,Z))-g(\nabla_XY,Z)-g(Y,\nabla_XZ)=0\\
{\text\rm ii)}&\quad T_\nabla=0.
\endaligned
\tag1
$$
}
\vskip 0.2cm

The Theorem 1 is a generalization of a well-known theorem from differential geometry. A proof can be found in [Go].

The conditions  (1) can be rewritten in the form:
$$
\aligned
{\text\rm i)}&\quad\nabla_XY=\nabla_YX+p_0[X,Y]\\
{\text\rm ii)}&\quad Z(g(X,Y))=g(\nabla_ZX,Y)+g(X,\nabla_ZY).
\endaligned
\tag2
$$
By cyclic permutation of  $X,Y,Z$ in (2 $ii)$) and by summation we get:
$$
\aligned
&g(\nabla_XY,Z)=\frac 12\{X(g(Y,Z))+Y(g(Z,X))-Z(g(X,Y))+\\
+&g(Z,p_0[X,Y])+g(Y,p_0[Z,X])-g(X,p_0[Y,Z]\}.
\endaligned
\tag3
$$
Let $q^i,\ (i=1,\dots,n)$ be local coordinates on $M$, such that the first $m$
coordinate   vector   fields  $\frac{\partial}{\partial q^i}$ are projected  by
projection  $p_0$ into vector fields $e_a,\ (a=1,\dots,m)$, generating the
distribution  $V$: $p_0(\frac{\partial}{\partial q^i})=p_i^a(q)e_a$. Vector
fields $e_a$ can be expressed in the basis $\frac{\partial}{\partial q^i}$ as
$e_a=B^i_a\frac{\partial}{\partial q^i}$, with $B^i_ap^b_i=
\delta_a^b$. Now we give  coordinate expressions for the   coefficients of the
connection
$\Gamma_{ab}^c$, defined as  $\nabla_{e_a}e_b=\Gamma_{ab}^ce_c$.
From (3) we get:
$$
\Gamma_{ab}^c=\left\{_{ab}^c\right\}+g_{ae}g^{cd}\Omega^e_{bd}+g_{be}g^{cd}
\Omega^e_{ad}-\Omega^c_{ab},
\tag4
$$
where $\Omega$ is obtained from $p_0[e_a,e_b]=-2\Omega_{ab}^ce_c$ as:
$$
2\Omega_{ab}^c=p_i^ce_a(B^i_b)-p_i^ce_b(B^i_a),
$$
and $\left\{_{ab}^c\right\}=\frac12g^{ce}(e_a(g_{be})+e_b(g_{ae})-e_e(g_{ab}))$.

It was shown in Section 1.3. that the equations of a nonholonomic
mechanical problem, without external forces, are  geodesic equations for the
connection  $\tilde\nabla$. The connection $\tilde\nabla$
is obtained by projection on the sub-bundle  $V$ of the Levi-Civita connection $\nabla$ for the
metric $g$. The question is: what is a relationship
between the connection   $\tilde\nabla$ and the metric $\tilde g$, induced   from   $g$ on $V$.
\vskip 0.2cm

{\bf Proposition 1.} {\it The connection  $\tilde\nabla$, obtained  by projecting metric
torsion-less connection $\nabla$ for the  metric $g$, is the metric torsion-less
connection for the  induced metric $\tilde g$ if
the projector $p_0$ is orthogonal.}
\vskip 0.2cm
{\bf Proof.} Let $p_0:TM\rightarrow V$ be the  orthogonal projector.

a) We need to prove $\tilde \nabla \tilde g =0$. For
arbitrary $X,Y,Z\in\Gamma(V)$ we have:
$$
\tilde{\nabla}_X \tilde g (Y,Z)=X(\tilde g (Y,Z))-\tilde g
(\tilde{\nabla}_XY,Z)-\tilde g (Y,\tilde{\nabla}_XZ).
\tag5
$$
Since $\tilde g$ is induced by $g$, it follows that $\tilde g (Y,Z)=g(Y,Z)$.
In the same way, $\tilde {\nabla}_XY=p_0\nabla_XY=\nabla_XY-U$, where
$U\in \Gamma(V^\perp)$ is a vector field projected with $p_0$ into $0$.
From the orthogonality condition, $U$ is orthogonal on $X,Y$ and $Z$ relatively
to the  metric  $g$, so we get: $\tilde g (\tilde{\nabla}_XY,Z)=
g (\tilde{\nabla}_XY,Z)=g(\nabla_XY-U,Z)=g(\nabla_XY,Z)$. Similarly, $\tilde g
(Y,\tilde{\nabla}_XZ)=g(Y,\nabla_XZ)$. Plugging into (5), we get:
$$
\tilde{\nabla}_X \tilde g (Y,Z)=\nabla_X g (Y,Z)\quad X,Y,Z \in \Gamma(V),
$$
and from the assumption $\nabla g=0$ we get $\tilde{\nabla} \tilde g=0$.

b) We need to show that the connection  $\tilde \nabla$ is torsion-less.
$$
\aligned
&T_{\tilde\nabla} (X,Y)=\tilde{\nabla}_XY-\tilde{\nabla}_YX-p_0[X,Y]\\
=&p_0\nabla_XY-p_0\nabla_YX-p_0[X,Y]=p_0(\nabla_XY-\nabla_YX-[X,Y]),
\endaligned
$$
and since $\nabla$ is free of torsion, the same is valid for $\tilde \nabla$.
\qed
\medskip

{\bf Note.} Both the Wagner and the  Shouten tensor, as we will see later,
depend
on the choice of the projector. Wagner defined  curvature tensor for a metric
which is defined on the  distribution $V$. If we start from some mechanical
 problem,
then there is a metric on the whole  $TM$, which is afterwards induced   on $V$. According to the last
Proposition, in order to get projected connection which is metric for the induced metric,
one is obliged to choose the orthogonal projector. That means, that for  mechanical systems there is a unique  choice
of a projector.
\medskip

The problem of definition of the curvature  tensor for  nonholonomic connections was considered for the first time by
Shouten. He  defined the curvature tensor in the following way:
\vskip 0.2cm

{\bf Definition 3.} {\it The Shouten tensor} is a mapping $K:\Gamma(V)\times
\Gamma(V)\times\Gamma(V)\rightarrow \Gamma(V)$ defined by:
$$
K(X,Y)Z=\nabla_X(\nabla_YZ)-\nabla_Y(\nabla_XZ)-\nabla_{p_0[X,Y]}Z-
p_0[q_0[X,Y],Z],
\tag6
$$
where $X, Y, Z\in \Gamma(V).$
\vskip 0.2cm

To check that the Definition 3 is correct, one has to verify that $K$ is of tensor nature,
i.e. that it is linear on $X,Y,Z$ relatively to the multiplication by smooth functions on $M$.
Really, by direct check [Go] we get:
$$
\aligned
K(fX,Y)Z&=fK(X,Y)Z,\\
K(X,Y)(fZ)&=fK(X,Y)Z,\\
K(X,Y)Z&=-K(Y,X)Z.
\endaligned
$$
In comparison to the curvature tensor for  connections on  $M$, we see that Shouten tensor
(6) has one term more, the last one in (6), and that in the third term $p_0$ appears.  The last term gives a
correction in order that  $K$ be a  tensor. Note that without that last term
linearity for $Z$ relatively to the multiplication by smooth functions would  not
be satisfied.

A mapping $K(X,Y):Z\rightarrow K(X,Y)Z$ is a morphism of $C^\infty(M)$-
module $\Gamma(V)$. Since $K$ is anti-symmetric relatively to $X,Y$, a $C^\infty(M)$-linear  mapping
$\Gamma(K):\Gamma(\wedge^2V)\rightarrow \Gamma(End(V,V))$ can be corresponded to the Shouten tensor by the condition:
$$
\Gamma(K)(X\land Y)Z=K(X,Y)Z,\quad X,Y,Z\in\Gamma(V),
$$
where $\wedge^2V$ is the space of bivectors.

\

\

\centerline{\bf \S 3. The Wagner tensor}

\

\centerline{\bf 3.1. The Wagner construction}

\

Wagner constructed a curvature tensor starting from the integrability condition for the tensor
equation $\nabla X=U$ where $U\in End(V,V),\ X\in \Gamma(V)$. If the curvature
tensor is zero, then  absolute parallelism should take place, i.e. a covariantly constant   vector    field in any direction should
exist     , which is
equivalent to the integrability of the equations $\nabla X=0$. Wagner noticed that if the degree of nonholonomicity is greater then 1, then the Shouten tensor does not
satisfy the condition of absolute parallelism, and he suggested a   correction.
The idea is the following. One starts with some metric $g$ on $V$.The metric $g$ is going to be extended  to each
sub-bundle     $V_i$ of the flag $V=V_0\subset V_1\subset\dots\subset V_N=TM$.
 The next step, the connection on $V_i$
and the curvature tensor analogous to the Shouten tensor are going
to be  defined. In this way, in the  $N$-th step, the curvature
tensor which satisfies the  absolute   parallelism condition is
constructed. The basic Wagner's paper where this was performed  is
[Wa1].

Let a metric $g$ be defined on  $k$-dimensional vector space $W$. Then a
metric $g^\land$ on $\wedge^2W$ is defined by the expression:
$$
g^\land(x_1\land y_1,x_2\land y_2)=
\left\vert\matrix g(x_1,x_2)&g(x_1,y_2)\\
g(y_1,x_2)&g(y_1,y_2)
\endmatrix\right\vert .
\tag1
$$
(The isomorphism $\varphi:\wedge^2W^*\rightarrow (\wedge^2W)^*$
$$
\varphi(f\land g)(x\land y)=\omega(x,y)=f(x)g(y)-f(y)g(x).
$$
is used here.)
\vskip 0.2cm

{\bf Lemma 1.} {\it If  $g$ is positively  definite  form on $W$, then
$g^\land$ is also positively  defined form  on $\wedge^2W$.}
\vskip 0.2cm

Consider a mapping
$$
\Delta :\wedge^2\Gamma(V)\rightarrow\Gamma(TM)/\Gamma(V),
$$
defined by
$$
\Delta(X\land Y)=[X,Y]\mod \Gamma(V)\ ,\quad X,Y\in\Gamma(V).
$$
The mapping $\Delta$ is $C^\infty(M)$ - linear:
$$
\aligned
\Delta(fX\land Y)&=[fX,Y]\mod\Gamma(V)=\left\{-Y(f)X+f[X,Y]\right\}
\mod\Gamma(V)=\\
&=f[X,Y]\mod\Gamma(V)=f\Delta(X\land Y).
\endaligned
$$
Observe that $\Im (\Delta)$ is not always equal to $\Gamma(TM)/\Gamma(V)$, but
it is its
 $C^\infty (M)$-submodule, and denote
$$
\Gamma(V_1)=\left\{X\in \Gamma(TM)\vert
X\mod\Gamma(V)\in Im(\Delta)\right\}.
$$
So, we get a sequence of $C^\infty$ submodules $\Gamma(V_0)\subset\dots\subset
\Gamma(V_N)=\Gamma(TM)$,  defined by:
$$
\aligned
&\Gamma(V_i)=\left\{ X\in \Gamma(TM)\vert X\mod\Gamma(V_{i-1})\in
\Im(\Delta_{i-1})\right\},\\
&\Delta_i(X\land Y)=[X,Y]\mod\Gamma(V_i),\quad i=1,\dots,N,
\endaligned
\tag2
$$
where  $V=V_0,\ \Delta=\Delta_0$.  Note that the sequence of
sub-bundles  $V_0\subset V_1\subset\dots \subset V_N=TM$ is a flag
of the distribution $V$, and  $N$ is the degree of
nonholonomicity, since we reduced our attention to the case of
regular distributions. The mapping
$\Delta_i:\wedge^2V_i\rightarrow TM/{V_i}$ is called {\it the
$i$-th tensor of nonholonomicity} of the  distribution $V$.

For every point $x\in M$, there is a  factor space $V_{i+1,x}/{V_{i,x}}$ with the projection $\pi_i:V_{i+1,x}
\rightarrow V_{i+1,x}/V_{i,x}$. Suppose  the mappings $\theta_{i,x}:V_{i+1,x}/V_{i,x}\rightarrow
R_{i,x}$ are defined, where  $R_{i,x}$ are some sub-spaces,
chosen transversely to $V_{x,i}$, so that $V_{i,x}\oplus R_{i,x}=V_{i+1,x}$.
Mappings $q_i=\theta_i\cdot \pi_i$ and  $p_i=1_{V_{i+1}}-q_i$ are  the projectors
onto $R_i$ and $V_i$ respectively.
Now we are going to extend the metric from  $V$ to the whole $TM$.
\vskip 0.2cm

{\bf Theorem 1.} {\it Let the distribution $V$ with metric $g$ and
mappings  $\theta_0,\dots\theta_{N-1}$ are given. Then there exists a unique metric
tensor $G$ on $TM$, which satisfies the conditions:
\item {1.} $G\vert_V=g$.
\item {2.} In the direct sum $TM=V_0\oplus R_0\oplus\dots\oplus R_{N-1}$
the components are mutually orthogonal.
\item {3.} $(G\vert_{R_i})^{-1}=\theta_i\cdot\Delta_i\cdot\left((G\vert_{V_i})
^\land\right)^{-1}\cdot(\theta_i\cdot\Delta_i)^*$.}
\vskip 0.2cm

{\bf Proof.}
For an arbitrary point $x$ on $M$ we have $T_xM=V_{0,x}\oplus
R_{0,x}\oplus...\oplus R_{N-1,x}$. Define
$G\vert_{R_i,x}=g_{i+1,x}$ by the condition  3 of this Theorem. By the
previous Lemma,  $g_{0,x}^\wedge$ is a positively defined             form  on
$\wedge^2 V_0$, so it is  $(g_{0,x}^\wedge)^{-1}$ on
$(\wedge^2 V_0)^*$. The operation of  conjugation preserves  positive
definitness, so we get that  $g_{1,x}$ is also a  positively
definite  form. By iterations we get that
$g_{i+1,x}$ are positively definite.
\qed
\vskip 0.2cm

Coordinate expressions for the metric enlarged from  $V_{i-1}$ to $V_{i}=V_{i-1}
\oplus R_{i-1}$ are obtained in the following way. Let the vectors $e_{a_{i-1}}$
span $V_{i-1}$.
Corresponding  dual base denote  by $e^{a_{i-1}}$. If
$X_{a_i}e^{a_i}$ is a given  1-form on $R_{i-1}$, then:
$$
\aligned
\overset i\to g(X_{a_{i}}e^{a_i})&=\overset i\to g^{a_ib_i}X_{a_i}e_{b_i}\\
&=(\theta_{i-1}\cdot\Delta_{i-1})(G_{V_{i-1}}^\wedge)^{-1}(\theta_{i-1}
\cdot\Delta_{i-1})^*
(X_{a_i}e^{a_i})\\
&=(\theta_{i-1}\cdot\Delta_{i-1})(G_{V_{i-1}}^\wedge)^{-1}(X_{a_i}
\overset i-1\to M_{a_{i-1}b_{i-1}}^{a_i}e^{a_{i-1}}\wedge e^{b_{i-1}})\\
&=(\theta_{i-1}\cdot\Delta_{i-1})(g^\wedge)^{a_{i-1}b_{i-1}c_{i-1}d_{i-1}}
(\overset i-1\to M_{a_{i-1}b_{i-1}}^{a_i}
X_{a_i}e_{c_{i-1}}\wedge e_{d_{i-1}})\\
&=(g^\wedge)^{a_{i-1}b_{i-1}c_{i-1}d_{i-1}}
(X_{a_i}
\overset i-1\to M_{a_{i-1}b_{i-1}}^{a_i}
\overset i-1 \to M_{c_{i-1}d_{i-1}}^{b_i}e_{b_i}),
\endaligned
$$
where $g^{\wedge a_{i-1}b_{i-1}c_{i-1}d_{i-1}}$ is the inverse metric tensor
for $g^\wedge$ defined by (1), and  $\overset i-1 \to M_{c_{i-1}d_{i-1}}^
{b_i}$ are coordinate  expressions for the $(i-1)$-th tensor of nonholonomicity $\Delta_{i-1}$.
It is obvious that
$$
g^{\wedge a_{i-1}
b_{i-1}c_{i-1}d_{i-1}}=\frac 12 (\overset i-1\to g^{a_{i-1}c_{i-1}}\
\overset i-1\to g^{b_{i-1}d_{i-1}}-\overset i-1\to g^{a_{i-1}d_{i-1}}\
\overset i-1\to g^{b_{i-1}c_{i-1}}),
$$
so, finally we get
$$
\overset i\to g^{a_ib_i}=\overset i-1\to M_{a_{i-1}b_{i-1}}^{a_i}
\overset i-1\to M_{c_{i-1}d_{i-1}}^{b_i}
\overset i-1\to g^{a_{i-1}c_{i-1}}\ \overset i-1\to g^{b_{i-1}d_{i-1}}.
$$

Let us define morphism of vector bundles $\mu_i:V_{i+1}\rightarrow
\wedge^2V_i$, by:
$$
\mu_i=\left(B_i^\land\right)^{-1}\cdot(\theta_i\cdot \Delta_i)^*\cdot G_{i+1}
\vert_{R_i}\cdot \theta_i\cdot \pi_i .
\tag3
$$
So, if  $X\in \Gamma(V_i)$, then $\mu_i(X)=0$.

Now we get coordinate expressions
for $\mu_i$:
$$
\aligned
\mu_{i-1}(e_{a_i}) &=\overset *\to M^{a_{i-1}b_{i-1}}_{a_i}e_{a_{i-1}}\wedge
e_{b_{i-1}} \\
&=\overset i-1\to M_{c_{i-1}d_{i-1}}^{b_i}\overset i\to g_{a_i
b_i}\ \overset i-1\to g^{c_{i-1} a_{i-1}}\ \overset i-1\to
g^{d_{i-1}b_{i-1}} e_{a_{i-1}}\wedge
e_{b_{i-1}}.\\
\endaligned
$$
Coordinate expressions for $\mu_i$ and those for metrics are in the agreement
with the original Wagner's paper [Wa1].

We are ready to expose Wagner's construction for the curvature tensor
for  nonholonomic systems.

Denote by $\overset 0\to\nabla$ the connection for the metric  $g_0$ on
$V_0$, and by $\overset 0 \to {K_\sq}$ the Shouten tensor. Define
$\overset 1 \to\sq :\Gamma(V_1)\times\Gamma(V_0)\rightarrow\Gamma(V_0)$ by:
$$
\overset 1\to{\sq}_XU=\overset 0\to {\nabla}_{p_0X}U+\overset
0\to{K}_\sq (\mu_0(X))(U)+p_0[q_0 X,U],
$$
and  $\overset 1\to {K}_\sq:\wedge^2V_1\rightarrow End(V_0)$ by the condition:
$$
\Gamma(\overset 1\to K_\sq)(X\land Y)(U)=\overset 1 \to \sq_X\overset1\to\sq_Y
U-\overset 1 \to \sq_Y\overset1\to\sq_X U-\overset 1\to\sq_{p_1[X,Y]}U-
p_0[q_1[X,Y],U],
$$
where $X,Y\in \Gamma(V_1),\ U\in\Gamma(V_0)$.

Similarly,  by  induction: $\overset i\to \sq :\Gamma(V_i)\times\Gamma(V_0)
\rightarrow\Gamma(V_0)$
$$
\aligned
&\overset i\to \sq_XU=\overset {i-1} \to \sq_{p_{i-1}X}U+\overset {i-1}\to
K_\sq(\mu_{i-1}(X))(U)+p_0[q_{i-1}X,U],\\
&\overset i\to K_\sq:\wedge^2V_i\rightarrow End(V_0)\quad X,Y\in \Gamma(V_i),
\ U\in\Gamma(V_0),\\
&\Gamma(\overset i\to K_\sq)(X\land Y)U=\overset i \to \sq_X\overset i\to\sq_Y
U-\overset i \to \sq_Y\overset i\to\sq_X U-\overset i\to\sq_{p_i[X,Y]}U-
p_0[q_i[X,Y],U].
\endaligned
$$
Finally for $i=N$ we get:
$$
\aligned
&\overset N\to \sq :\Gamma(V_N)\times\Gamma(V_0)
\rightarrow\Gamma(V_0),\\
&\overset N\to \sq_XU=\overset {N-1} \to
\nabla_{p_{N-1}X}U+\overset {N-1}\to
K_\sq(\mu_{N-1}(X))(U)+p_0[q_{N-1}X,U],
\endaligned
\tag{4}
$$
$$
\aligned
&\overset N\to K_\sq:\wedge^2V_N\rightarrow End(V_0),\quad X,Y\in \Gamma(V_N),
\ U\in\Gamma(V_0),\\
&\Gamma(\overset N\to K_\sq)(X\land Y)U=\overset N \to \sq_X\overset N\to\sq_Y
U-\overset N \to \sq_Y\overset N\to\sq_X U-\overset N\to\sq_{[X,Y]}U ,
\endaligned
\tag{5}
$$
because $p_N=id$, and $q_N=0$.
\vskip 0.2cm

{\bf Theorem 2.} {\it  Mappings $\overset i \to \sq$, satisfy the
following conditions:}
$$
\aligned
&1.\quad \overset i\to \sq_{fX+gY}U=f\overset i\to \sq_XU+g\overset
i\to\sq_YU,\quad f,g\in C^\infty(M)\\
&2.\quad \overset i\to \sq_X(fU)=X(f)U+f\overset i\to \sq_XU,\quad X,Y\in
\Gamma(V_i)\\
&3.\quad \overset N \to \sq \quad{\text {\it is a linear connection on
the vector bundle}}\; V.
\endaligned
$$
\medskip
The proof follows by direct calculations.
\vskip 0.2cm

Since   $\overset N\to\sq$ is a connection on the vector bundle, according to
the Theorem 2, we get that    $\overset N\to K_\sq$ is the curvature tensor of the  vector
bundle     $V$ over $M$, relative to the connection $\overset N\to\sq$, and it is
called {\it the Wagner tensor } of nonholonomic manifold.
\medskip

{\bf Note.} In  [Go], the Wagner tensor is defined in a slightly
different manner, as a mapping
$K_\sq:\wedge^2\Gamma(V_N)\rightarrow\Gamma(End(V_{N-1}))$. The
way  presented here is in  agreement with the original   Wagner
paper [Wa1], as it is going to be clear from the coordinate
expressions given below.

\

\

\centerline{\bf  3.2. Coordinate expressions for the  Wagner tensor}

\

Now we are going to derive the  coordinate expressions for the
Shouten tensor and the Wagner tensor. The Latin indices $a_i$ run
in the intervals $1,...,n_i$, where  $n_i=dimV_i$, and  Greek
indices $\alpha$ in the interval $1,\dots,n$. Let $e_a$ be vector
fields spanning the distribution $V$, and $p_0$ and $q_0$ the
projectors to $V$ and  $V^\perp$ respectively. The components of
the
 Shouten
tensor $K_{abc}^d$ are derived from:
$$
K(e_a,e_b)(e_c)=K_{abc}^de_d.
$$
Plugging into (2.6) and using the properties of the  connection   $\nabla$ we get:
$$
K_{abc}^d=e_a(\Gamma_{bc}^d)-e_b(\Gamma_{ac}^d)+\Gamma_{ae}^d\Gamma_{bc}^e-
\Gamma_{be}^d\Gamma_{ac}^e+2\Omega_{ab}^e\Gamma_{ec}^d-M_{ab}^p\Lambda_{pc}^d.
\tag{6}
$$
Coefficients $\Lambda_{pc}^d$ are defined by $p_0[e_p,e_c]=\Lambda_{pc}^de_d,\
p=m+1,\dots,n$ and $M_{ab}^p$ are the components of the tensor of  nonholonomicity $\Delta$
defined by $M_{ab}^pe_p=q_0[e_a,e_b]$. Expressing $e_a$ in the basis of
coordinate vector fields  $\frac{\partial}{\partial q^i}$ as
$e_a=B_a^i\frac{\partial}{\partial q^i}$ and plugging into (6), we get coordinate
expressions for the  Shouten tensor, which coincide with those obtained in [Wa1].

Denote by $\overset i \to \Pi_{a_ib}^c$ the components of the connection for
$\overset i\to \sq$ defined by    $\overset i\to \sq_{e_{a_i}}e_b=\overset i
\to \Pi_{a_ib}^ce_c$, where the vector fields  $e_{a_i}$ span the
distribution $V_i$. So, we get:
$$
\overset i\to\Pi_{a_ib}^c=\overset {i-1}\to
p_{a_i}^{a_{i-1}}\overset {i-1}\to \Pi_{a_{i-1}b}^c+\overset *\to
M_{a_i}^{a_{i-1}b_{i-1}}\overset {i-1} \to
K_{a_{i-1}b_{i-1}b}^c+\overset {i-1} \to q_{a_i}^p\Lambda_{pb}^c
\tag{7}
$$
In the same way we get coordinate expressions for $\overset i\to K_\sq$:
$$
\overset i\to K_{a_ib_ic}^d=e_{a_i}(\overset i \to \Pi_{b_ic}^d)-
e_{b_i}(\overset i \to \Pi_{a_ic}^d)+\overset i \to \Pi_{a_ie}^d
\overset i \to \Pi_{b_ic}^e-\overset i \to \Pi_{b_ie}^d
\overset i \to \Pi_{a_ic}^e
+2\overset i\to\Omega_{a_ib_i}^{c_i}
\overset i \to \Pi_{c_ic}^d-\overset i\to M_{a_ib_i}^p\Lambda_{pc}^d.
\tag{8}
$$
$\overset i \to p$ and  $\overset i \to q$ are the corresponding
projectors to $V_i$ and  $V_i^\perp$ and $\overset i \to
\Omega_{a_i,b_i}^{c_i}$ is defined by $2\overset i\to \Omega_{a_i,b_i}^
{c_i}e_{c_i}=-\overset i \to p[e_{a_i},e_{b_i}]$, while $\overset i\to
M_{a_i b_i}^p$ are the components of the  $i$-th tensor of nonholonomicity, defined by (2).

Finally, for $i=N$, we get coordinate expressions for the Wagner tensor
$$
\overset N\to K_{a_Nb_Nc}^d=e_{a_N}(\overset N \to \Pi_{b_Nc}^d)-
e_{b_N}(\overset N \to \Pi_{a_Nc}^d)+\overset N \to \Pi_{a_Ne}^d
\overset N \to \Pi_{b_Nc}^e-\overset N \to \Pi_{b_Ne}^d
\overset N \to \Pi_{a_Nc}^e
+2\overset N\to\Omega_{a_Nb_N}^{c_N}\overset N\to\Pi_{c_Nc}^d .
\tag{9}
$$
The vector fields  $e_{a_N}$ are now spanning the whole $TM$.

\

\

\centerline{\bf 3.3. Absolute parallelism and the Wagner tensor}

\

We start from the equation
$$
\nabla W=U\ ,\quad U\in\Gamma(End(V)),\ W\in\Gamma(V).
\tag{10}
$$
The question is if for a given endomorphism $U$ and for every $X\in
\Gamma(V)$, the equation:
$$
\nabla_XW=U_X
$$
has a solution. From (10) we get:
$$
\aligned
&\nabla_X\nabla_YW-\nabla_Y\nabla_XW-\nabla_{p_0[X,Y]}W-p_0[q_0[X,Y],W]=\\
=&\nabla_XU_Y-\nabla_YU_X-U_{p_0[X,Y]}-p_0[q_0[X,Y],W].
\endaligned
$$
So, there exists $X\in \Gamma(V_1)$ such that:
$$
\overset 0\to K(\mu_0(X))(W)+p_0[q_0X,W]=U^\nabla(\mu_0(X)),
$$
where $U^\nabla(\mu_0(X))=\nabla_XU_Y-\nabla_YU_X-U_{p_0[X,Y]}$. Then:
$$
\nabla_{p_0X}W+\overset 0\to K(\mu_0(X))(W)+p_0[q_0X,W]=\overset 1\to U_X=
U^\nabla(\mu_0(X))+U_{p_0X}.
$$
The integrability conditions for the equation (10) are reduced to:
$$
\overset 1\to \sq W=\overset 1\to U.
\tag{11}
$$
In the same way, iteratively,  we reduce the  integrability condition for the
equation (10) to the condition:
$$
\overset i \to \sq W=\overset i\to U.
$$
Finally, for $i=N$ we get:
$$
\overset N \to \sq W=\overset N\to U.
$$
So :
$$
\overset N\to K(X\wedge Y)(W)=\overset N \to\sq_X\overset N \to\sq_YW-
\overset N \to\sq_Y\overset N \to\sq_XW-\overset N \to\sq_{[X,Y]}W
=\overset N \to\sq_X\overset N \to U_Y-\overset N \to\sq_Y\overset N \to U_X-
\overset N \to U_{[X,Y]}.
\tag{12}
$$
This equation  is the integrability condition for the equation  (10).
Therefore, in the case $U=0$, the necessary and sufficient condition for the
existence of the  vector
fields parallel along any direction is that  the Wagner tensor is equal to zero.

\

\

\centerline {\bf \S 4. The rolling disc}

\bigskip

\

Now, we are going to illustrate the theory exposed before by calculating the Wagner
tensors in two mechanical problems. In this section, we deal with a
homogeneous disc of the unit mass and radius $R$ rolling without
sliding on a horizontal plane.

Note that we are going to present only basic steps of the calculations.
As it is well known, the configuration space is $M=R^2\times SO(3)$. For local
coordinates we chose  $x$ and $y$ as  coordinates of the mass  center of the disc, and the Euler angles $\varphi,\psi, \theta$. Nonholonomic constraints
follow from the condition that the velocity of the point of contact of the disc and the plane should be equal to zero.
The two nonholonomic constraints are:
$$
\aligned
&\dot x+R\cos\varphi\dot \psi +R\cos \theta\cos\varphi\dot \varphi - R\sin
\theta\sin\varphi\dot\theta=0,\\
&\dot y+R\sin\varphi\dot \psi +R\cos \theta\sin\varphi\dot \varphi + R\sin
\theta\cos\varphi\dot\theta=0.
\endaligned
$$
Corresponding $1$-forms which define the three-dimensional distribution $V$ are:
$$
\aligned
\omega_1=&dx+R\cos\varphi d\psi +R\cos \theta\cos\varphi d\varphi - R\sin
\theta\sin\varphi d\theta,\\
\omega_2=&d y+R\sin\varphi d\psi +R\cos \theta\sin\varphi d\varphi + R\sin
\theta\cos\varphi d\theta.
\endaligned
$$
The vector fields which span the differential  system $N(V)$ are:
$$
\aligned
e_1&=R\cos\varphi\frac{\partial}{\partial x}+R\sin\varphi\frac{\partial}
{\partial y}-\frac{\partial}{\partial \psi},\\
e_2&=\cos\theta\frac{\partial}{\partial \psi}-\frac{\partial}{\partial
\varphi},\\
e_3&=R\sin\theta\sin\varphi\frac{\partial}{\partial x}-R\sin\theta\cos\varphi
\frac{\partial}{\partial y}+\frac{\partial}{\partial \theta}.
\endaligned
$$
First, let us calculate the degree of nonholonomicity of this mechanical system:
$$
\aligned
&[e_1,e_2]=-R\sin\varphi\frac{\partial}{\partial x}+R\cos\varphi
\frac{\partial}{\partial y}=T,\\
&[e_1,e_3]=0,\\
&[e_2,e_3]=-\sin\theta e_1.
\endaligned
$$
So, the  distribution $V$ is nonintegrable, and the whole $TM$ is not generated in the first step. From:
$$
\aligned
&[e_1,e_2]=T,\quad [e_1,e_3]=0,\quad [e_2,e_3]=-\sin\theta e_1,\\
&[e_1,T]=0,\quad [e_2, T]=R\cos\varphi\frac{\partial}{\partial x}+
R\sin\varphi\frac{\partial}{\partial y}=U,
\endaligned
$$
since  $e_1, e_2, e_3, T, U$ span the  tangent space in every point
of $M$, the degree of nonholonomicity is $2$.

It is well known that the  kinetic energy of the  system is:
$$
2T={\dot x}^2+{\dot y}^2+(A\sin^2\theta+C\cos^2\theta){\dot \varphi}^2+2
C\cos\theta\dot \varphi\dot \psi+C{\dot \psi}^2+(A+R^2\cos^2\theta){\dot
\theta}^2
$$
where  $A$ and $C$ are the principle  central moments of inertia of the  disc in the moving frame.
This gives a  metric on $M$:
$$
(g_{ij})=\left(\matrix 1&0&0&0&0\\
 0&1&0&0&0\\
0&0&A\sin^2\theta+C\cos^2\theta&C\cos\theta&0\\
0&0&C\cos\theta&C&0\\
0&0&0&0&A+R^2\cos^2\theta
\endmatrix\right).
$$
As it was pointed out after the Proposition 2.1, in mechanical problems we chose the orthogonal projector $p_0$ from $TM$ onto
$V$. The vector fields annulated by  $p_0$ are:
$$
\aligned
e_4&=-\sin\varphi(A+R^2\cos^2\theta)\frac{\partial}{\partial x}+
\cos\varphi(A+R^2\cos^2\theta)\frac{\partial}{\partial y}+
R\sin\theta\frac{\partial}{\partial \theta},\\
e_5&=C\cos\varphi\frac{\partial}{\partial x}+C\sin\varphi \frac{\partial}
{\partial y}+R\frac{\partial}{\partial \psi}.
\endaligned
$$
The vector fields  $e_a$ are expressed in the  basis $\frac{\partial}{\partial x^i}$ by
$e_a=B_a^i\frac{\partial}{\partial x_i}$. So we get:
$$
(B_a^i)=\left(\matrix R\cos\varphi&R\sin\varphi&0&-1&0\\
0&0&-1&\cos\theta&0\\
R\sin\theta\sin\varphi&-R\sin\theta\cos\varphi&0&0&1
\endmatrix\right).
$$
 From $p_0(\frac{\partial}{\partial x^i})=
p_i^ae_a$, we get the coordinates of the  projector:
$$
(p_i^a)=\left(\matrix \frac{R\cos\varphi}{C+R^2}&0&\frac{R\sin\theta\sin
\varphi}{A+R^2}\\
 & & \\
\frac{R\sin\varphi}{C+R^2}&0&\frac{-R\sin\theta\cos
\varphi}{A+R^2}\\
 & & \\
\frac{-C\cos\theta}{C+R^2}&-1&0\\
 & & \\
\frac{-C}{C+R^2}&0&0\\
 & & \\
0&0&\frac{A+R^2\cos^2\theta}{A+R^2}
\endmatrix\right).
$$
Similarly, for $q_0$ we get:
$$
(q_i^p)=\left(\matrix \frac{-\sin\varphi}{A+R^2}&\frac{\cos\varphi}{C+R^2}\\
 & \\
\frac{\cos\varphi}{A+R^2}&\frac{\sin\varphi}{C+R^2}\\
 & \\
0&\frac{R\cos\theta}{C+R^2}\\
 & \\
0&\frac R{C+R^2}\\
 & \\
\frac{R\sin\theta}{A+R^2}&0
\endmatrix\right).
$$
The induced metric $g_{ab}$ on $V$, is derived from $g_{ij}$:
$$
(g_{ab})=\left(\matrix R^2+C&0&0\\
0&A\sin^2\theta&0\\
0&0&A+R^2
\endmatrix\right).
$$
Now we calculate the components of the connection $\Gamma_{ab}^c$ for metric connection using coordinate expressions  (2.4).  We start with determining $\left\{_{ab}^c\right\}$. The only nonzero coefficients are:
$$
\left\{_{23}^2\right\}=\left\{_{32}^2\right\}=\frac{\cos\theta}{\sin
\theta},\quad \left\{_{22}^3\right\}=\frac{-A\sin\theta\cos\theta}{A+R^2}.
$$
The coefficients $\Omega$ we derive from $-2\Omega_{ab}^c=p_0[e_a,e_b]$. Having the expressions for the commutators of  $e_a$, it can easily be seen  that nonzero elements are:
$$
\Omega_{12}^3=-\Omega_{21}^3=\frac{R^2\sin\theta}{2(A+R^2)}, \quad
\Omega_{23}^1=-\Omega_{32}^1=\frac{\sin\theta}2 .
$$
From  (2.4) we get the following nonzero  components of the connection:
$$
\aligned
&\Gamma_{23}^1=\frac{-(2R^2+C)\sin\theta}{2(C+R^2)},\quad \Gamma_{32}^1=
\frac{C\sin\theta}{2(C+R^2)},\quad \Gamma_{23}^2=\Gamma_{32}^2=\frac{\cos
\theta}{\sin\theta},\\
&\Gamma_{13}^2=\Gamma_{31}^2=\frac{-C}{2A\sin\theta},
\quad \Gamma_{12}^3=\frac{C\sin\theta}{2(A+R^2)},\\
&\Gamma_{21}^3=\frac
{(2R^2+C)\sin\theta}{2(A+R^2)},\quad
\Gamma_{22}^3=\frac{-A\sin\theta\cos\theta}{A+R^2}.
\endaligned
$$

In order to get the components of the Shouten tensor  (see (3.6)), we are calculating  the  coefficients
$\Lambda$. From:
$$
[e_4,e_1]=0,\quad [e_4,e_2]=-\cos\varphi (A+R^2\cos^2\theta)\frac{\partial}
{\partial x}-\sin\varphi(A+R^2\cos^2\theta)\frac{\partial}{\partial y}-
R\sin^2\theta\frac{\partial}{\partial \psi},
$$
$$
[e_4,e_3]=-R^2\sin\varphi\cos\theta\sin\theta \frac{\partial}
{\partial x}+R^2\cos\varphi\cos\theta\sin\theta\frac{\partial}{\partial y}-
R\cos\theta\frac{\partial}{\partial \theta},\quad
[e_5,e_1]=0,
$$
$$
[e_5,e_2]=-C\sin\varphi\frac{\partial}{\partial x}
+C\cos\varphi\frac{\partial}{\partial y},\quad [e_5,e_3]=0,
$$
we get:
$$
\Lambda_{42}^1=\frac{-R(A+R^2\cos^2\theta-C\sin^2\theta)}{C+R^2},\quad
\Lambda_{43}^3=-R\cos\theta,\quad \Lambda_{52}^3=\frac{-RC\sin\theta}{A+R^2}.
$$
Similarly, for the  components of the  tensor of nonholonomicity we get:
$$
\overset 0\to M_{12}^4=\frac{R}{A+R^2},\quad \overset 1\to
M_{24}^5=\frac{A+R^2}{C+R^2},
$$
where the  projectors $p_1$ and $q_1$ to $V_1$ and $V_1^\perp$ are used.
Here $V_1$ is generated by the vector fields $e_1, e_2, e_3, e_4$:
$$
(p_i^a)=\left(\matrix \frac{R\cos\varphi}{C+R^2}&0&\frac{R\sin\theta\sin
\varphi}{A+R^2}&\frac{-\sin\varphi}{A+R^2}\\
 & & & \\
\frac{R\sin\varphi}{C+R^2}&0&\frac{-R\sin\theta\cos
\varphi}{A+R^2}&\frac{\cos\varphi}{A+R^2}\\
 & & & \\
\frac{-C\cos\theta}{C+R^2}&-1&0&0\\
 & & & \\
\frac{-C}{C+R^2}&0&0&0\\
 & & & \\
0&0&\frac{A+R^2\cos^2\theta}{A+R^2}&\frac{R\sin\theta}{A+R^2}
\endmatrix\right),\quad
(q_i^p)=\left(\matrix \frac{\cos\varphi}{C+R^2}\\
  \\
\frac{\sin\varphi}{C+R^2}\\
\\
\frac{R\cos\theta}{C+R^2}\\
 \\
\frac{R}{C+R^2}\\
 \\
0
\endmatrix\right).
$$
Expansion of the metric from  $V_0$ to $V_1$ is obtained from the coordinate
expression:
$\overset i\to g^{a_1b_1}=M_{ab}^{a_1}M_{cd}^{b_1}g^{ac}\ g^{bd}$ as:
$$
\aligned
&g^{44}=\frac{2R^2}{(A+R^2)^2(C+R^2)A\sin^2\theta},\\
&g_{44}=\frac{1}{g^{44}}.
\endaligned
$$
Similarly, we get the coordinate expressions for the metric expanded on  $V_2=TM$ by:
$$
\aligned
&g^{55}=\frac{4R^2}{A^2(C+R^2)^3\sin^4\theta},\\
&g_{55}=\frac{1}{g^{55}}.
\endaligned
$$
From the expanded  metric, as it was mentioned before, we get the components
for the
morphisms $\mu_0$ i $\mu_1$:
$$
\overset *\to M_4^{12}=(\overset 0\to M_{12}^4)^2g^{11}\
g^{22}=\frac{A+R^2} {2R},\quad \overset*\to
M_5^{24}=\frac{C+R^2}{2(A+R^2)}.
$$
Everything is prepared for calculation of the Wagner tensor. In the coordinate expressions for the Wagner
tensor, the first two indices take values from 1 to 5, and the second two from 1 to 3. from the antisymmetry for the first two indexes, there are  90
independent components of the  Wagner tensor. We are going to calculate
three components. All calculations are performed in three steps: the first step is the    Shouten tensor,
then the tensor $\overset 1\to K$ on $V_1$, and finally the Wagner tensor.
We are calculating only the necessary components.

We calculate the  component $K_{451}^2$ of the Wagner tensor.
$$
K_{451}^2 =e_4(\overset 2\to\Pi_{51}^2)-e_5(\overset 2\to\Pi_{41}^2)+
\overset 2\to\Pi_{4c}^2\overset 2\to \Pi_{51}^c-
\overset 2\to\Pi_{5c}^2\overset 2\to \Pi_{41}^c,
$$
$$
\overset 2\to \Pi_{51}^c=\overset *\to M_5^{24}\overset 1\to K_{241}^c,\quad
\overset 2\to \Pi_{5c}^2=\overset *\to M_5^{24}\overset 1\to K_{24c}^2,
$$
$$
\overset 2\to \Pi_{41}^c=
\overset 1\to \Pi_{41}^c=\overset *\to M_4^{12}\overset 0\to K_{121}^c,
$$
$$
\overset 2\to \Pi_{4c}^2=
\overset 1\to \Pi_{4c}^2=\overset *\to M_4^{12}\overset 0\to K_{12c}^2,
$$
$$
\overset 1\to K_{241}^c =e_2(\overset 1\to\Pi_{41}^c)-e_4(\Gamma_{21}^c)+
\Gamma_{2d}^c\overset 1\to \Pi_{41}^d-
\overset 1\to\Pi_{43}^c\Gamma_{21}^3,
$$
$$
\overset 1\to K_{24c}^2 =e_2(\overset 1\to\Pi_{4c}^2)-e_4(\Gamma_{1c}^2)+
\Gamma_{23}^2\overset 1\to \Pi_{4c}^3-
\overset 1\to\Pi_{4d}^2\Gamma_{2c}^d+2\overset 1\to\Omega_{24}^1\Gamma_{1c}^2,
$$
$$
\overset 1\to \Pi_{43}^c=
\overset *\to M_4^{12}\overset 0\to K_{123}^c+\Lambda_{43}^c,\quad
\overset 1\to \Pi_{4c}^3=
\overset *\to M_4^{12}\overset 0\to K_{12c}^3.
$$
So, for the component $K_{451}^2$, we need first the
coordinate expressions for the components $\overset 0\to K_{12c}^d$ of the Shouten tensor.
From  (3.6) we get:
$$
\overset 0\to K_{121}^1=0,\quad \overset 0\to K_{121}^2=\frac{-C(4R^2+C)}
{4A(A+R^2)},
$$
$$
\overset 0\to K_{121}^3=0,\quad \overset 0\to K_{122}^1=\frac{4R^2A+4R^4\cos^2
\theta+C^2\sin^2\theta}{4(A+R^2)(C+R^2)},
$$
$$
\overset 0\to K_{122}^2=\frac{R^2\cos \theta} {A+R^2},\quad \overset
0\to K_{122}^3=0,
$$
$$
\overset 0\to K_{123}^1=0,\quad \overset 0\to K_{123}^2=0,\quad
\overset 0\to K_{123}^3=\frac{R^2\cos\theta}{A+R^2}.
$$
Similarly, we get:
$$
\overset 1\to \Pi_{41}^1=0,\quad \overset 1\to \Pi_{41}^3=0,\quad
\overset 1\to \Pi_{41}^2=\frac{-C(4R^2+C)}{4AR},
$$
$$
\overset 1\to \Pi_{42}^2=R\cos\theta,\quad \overset 1\to
\Pi_{42}^3=0,
$$
$$
\overset 1\to \Pi_{43}^1=0,\quad\overset 1\to \Pi_{43}^2=0,\quad
\overset 1\to \Pi_{43}^3=0.
$$
Therefore:
$$
\overset 1\to K_{241}^2=0,\quad \overset 1\to K_{242}^2=0,\quad
\overset 1\to K_{241}^1=0,
$$
$$
\overset 1\to
K_{243}^2=\frac{8R^4A\sin^2\theta-10R^2C^2\sin^2\theta-
C^3\sin^2\theta+8R^2AC\sin^2\theta+4R^2AC-8R^4C\sin^2\theta+4R^4C\cos^2\theta}
{8AR\sin\theta(C+R^2)}
$$
So
$$
\overset 2\to\Pi_{51}^1=\overset *\to M_5^{24}\overset 1\to K_{241}^1=0,\quad
\overset 2\to\Pi_{51}^2=\overset *\to M_5^{24}\overset 1\to K_{241}^2=0,\quad
\overset 2\to\Pi_{52}^2=\overset *\to M_5^{24}\overset 1\to K_{242}^2=0.
$$
Finally, we get
$$
K_{451}^2=0 .
$$

In the same way, we can calculate the other components of the  Wagner tensor.
For example, we are calculating also  $K_{121}^2$ and $K_{121}^3$.

From
$$
K_{121}^2=e_1(\overset 2\to\Pi_{21}^2)-e_2(\overset 2\to\Pi_{11}^2)+
\overset 2\to\Pi_{1c}^2\overset 2\to\Pi_{21}^c-
\overset 2\to\Pi_{2c}^2\overset 2\to\Pi_{11}^c+2\overset 2\to\Omega_{12}^{a_2}
\overset 2\to\Pi_{{a_2}1}^2,
$$
we get:
$$
K_{121}^2=\Gamma_{1c}^2\Gamma_{21}^c+2\overset 2\to\Omega_{12}^{a_2}
\overset 2\to\Pi_{{a_2}1}^2,
$$
and finally:
$$
K_{121}^2=0.
$$
Similarly $K_{133}^1=\frac{C^2}{4A(R^2+C)}$.

\

\centerline {\bf \S 5. Ball rolling on the fixed sphere}

\

Now we will give a construction of Wagner tensor for the system of
a homogeneous ball of unit mass o  rolling on the fixed sphere
$S^2$. Denote the diameters of the ball and the sphere by $r_2,
r_1$ respectively. This system has five degrees of freedom. Let us
introduce the following coordinates: the spherical coordinates
$\alpha, \beta$ on  $S^2$ and the Euler angles $\psi, \varphi,
\theta$ which determine  position of the ball. Nonholonomic
constraints are derived from the condition that velocity of the
contact point is equal to zero. There are two independent
nonholonomic constraints:
$$
\aligned
(1+k)&\dot \beta +\sin (\psi-\alpha) \dot \theta- \sin\theta \cos(\psi-\alpha)
\dot\varphi=0\\
(1+k)&\dot\alpha+\tan\beta\cos(\psi-\alpha)\dot\theta
+[\tan\beta\sin\theta\sin(\psi-\alpha)-\cos\theta]\dot\varphi-\dot\psi=0,
\endaligned
$$
where $k=r_1/r_2$. So, we assume $r_2=1$. Corresponding  1-forms that 
define the three-dimensional distribution $V$ are:
$$
\aligned
\omega_1=(1+k)&d \beta +\sin (\psi-\alpha) d \theta- \sin\theta \cos(\psi-\alpha)
d\varphi\\
\omega_2=(1+k)&d\alpha+\tan\beta\cos(\psi-\alpha)d\theta
+[\tan\beta\sin\theta\sin(\psi-\alpha)-\cos\theta]d\varphi-d\psi=0.
\endaligned
$$

Vector fields:
$$
\aligned
X_1&=\frac{\partial}{\partial\alpha}+(1+k)\frac{\partial}{\partial\psi}\\
X_2&=\tan\beta\sin\theta\frac{\partial}{\partial\alpha}-
(1+k)\sin \theta\cos(\psi-\alpha)\frac{\partial}{\partial \theta}-
(1+k)\sin(\psi-\alpha)\frac{\partial}{\partial \varphi} \\
&+ (1+k)\cos\theta\sin(\psi-\alpha)\frac{\partial}{\partial\psi}\\
X_3&=\sin\theta\frac{\partial}{\partial\beta}-
(1+k)\sin \theta\sin(\psi-\alpha)\frac{\partial}{\partial \theta}-
(1+k)\cos(\psi-\alpha)\frac{\partial}{\partial \varphi}\\
&- (1+k)\cos\theta\cos(\psi-\alpha)\frac{\partial}{\partial\psi}\\
\endaligned
$$
span the differential system $N(V)$.
Since
$$
\aligned
[X_1, X_2]&=(0, 0, k\,{\cos}\theta\,{\cos}(\psi - \alpha)\,(1
 + k), - k\,{\cos}(\psi - \alpha)\,(1 + k),\\
 & k\,{\sin}\theta
\,{\sin}(\psi - \alpha)\,(1+ k))\\
[X_1, X_3]&=(0, 0, k\,{\cos}\theta\,{\sin}(\psi - \alpha)\,(1
 + k), - k\,{\sin}(\psi - \alpha)\,(1 + k),\\
&- k\,{\sin}\theta\,{\cos}(\psi - \alpha)\,(
1 + k))\\
[X_2, X_3]&=(\frac{- \sin^2\theta + (1+k){\sin}\theta\,
{\sin}(\psi - \alpha)\,
{\cos}\theta\,{\sin}\beta\,{\cos}\beta}{\cos^2\beta},\\
&- {sin}\theta\,{\cos}(\psi - \alpha){\cos}\theta(1+k),\\
& \frac{ (1 + k)^2( 2\,{\cos^2}\theta\,{\cos}\beta\,
-{\cos}\beta)-(1+k){\sin}\theta
\,{\sin}(\psi - \alpha)\,{\sin}\beta\,{\cos}\theta)}{{\cos}\beta},\\
&
-\frac {(1 + k)^2\cos\theta\,\cos\beta\ -(1+k){\sin}\beta\,{\sin}\theta\,
{\sin}(\psi - \alpha)}{\cos\beta}, \\
&\frac {{\sin}\beta\,
{\cos}(\psi - \alpha)\,\sin^2\theta\,(1 + k)}
{{cos}\beta})
\endaligned
$$
 the degree of nonholonomicity is equal to one.

From the kinetic energy of the system:
$$
2T=(1+k)^2(\dot \beta^2+\cos^2\beta\dot\alpha^2)+A(\dot\psi^2+\dot\varphi^2+
\dot\theta^2+2\cos\theta\dot\varphi\dot\psi),
$$
where $A$ is the inertia momentum of the ball, the formula for the  metric is derived
$$
(g_{ij})=\left(\matrix
            (1+k)^2\cos^2\beta&0&0&0&0\\
            0&(1+k)^2&0&0&0\\
            0&0&A&A\cos\theta&0\\
            0&0&A\cos\theta&A&0\\
            0&0&0&0&A\endmatrix\right).
$$
We  choose the orthogonal projector $p_0$.
The vector fields orthogonal to the distribution $V$ are:
$$
\aligned
X_4&=A \cos(\psi-\alpha)\frac{\partial}{\partial\alpha}+
A\tan\beta\cos^2\beta\sin(\psi-\alpha)\frac{\partial}{\partial\beta}\\
&-(1+k)\cos^2\beta\cos(\psi-\alpha)\frac{\partial}{\partial\psi}
+(1+k)\tan\beta\cos^2\beta\frac{\partial}{\partial\theta}\\
X_5&=A \sin\theta\frac{\partial}{\partial\beta}+
(1+k)\cos\theta\cos(\psi-\alpha)\frac{\partial}{\partial\psi}\\
&-(1+k)\cos(\psi-\alpha)\frac{\partial}{\partial\varphi}
+(1+k)\sin\theta\sin(\psi-\alpha)\frac{\partial}{\partial\theta}.
\endaligned
$$
So the induced metric on the distribution
$V$ is
$$
(g_{ab})=\left(\matrix
            A+\cos^2\beta&\sin\beta\cos\beta\sin\theta&0\\
            \sin\beta\cos\beta\sin\theta&\sin^2\theta(A+\sin^2\beta)&0\\
            0&0&\sin^2\theta(1+A)
            \endmatrix\right)
$$
Using formula (2.4) we get:
$$
\aligned
\Gamma_{11}^3&=\frac {{\sin}\beta\,{\cos}\beta}{{\sin}\theta\,(
1 + A)}\, ,
\Gamma_{12}^3=-\frac {1}{2} \,\frac {A\,k - A
 - 2 + 2\,\cos^2\beta}{1 + A},\\
\Gamma_{13}^1&=-\frac {(1 + k)\,
{\sin}\theta\,{\sin}\beta\,{\cos}\beta}{(1 + A)},\ \
\Gamma_{13}^2=\frac {1}{2} \,\frac {A\,k - A +
{\cos^2}\beta\,k - 2 + {\cos^2}\beta}{1 + A},\\
\Gamma_{21}^3&= \frac {1}{2} \,\frac {A + A\,k + 2
 - 2\,{\cos^2}\beta}{1 + A},\ \
\Gamma_{22}^2= -(1+k) {\cos}\theta\,{\cos}(\psi- \alpha),\\
\Gamma_{22}^3&=\frac {(A + \sin^2\beta)\,{\sin}\theta\,
{\sin}\beta}{{\cos}\beta\,(1 + A)},\\
\Gamma_{23}^1&= \frac {k+1}{2}
\frac{-A\sin^2\theta+\cos^2\beta-1+\cos^2\beta\cos^2\theta+\cos^2\theta}{1+A},\\
\Gamma_{23}^2&=- \,\frac {(2\,A -
(1+k){\cos^2}\beta + 2)\,{\sin}\theta\,{\sin}\beta}{2{\cos}\beta\,(1 + A)}, \\
\Gamma_{23}^3&= -(1+k) {\cos}\theta\,{\cos}(\psi-\alpha),\ \
\Gamma_{31}^1=\frac {1}{2} \,\frac {( - 1 + k)\,
{\cos}\beta\,{\sin}\beta\,{\sin}\theta}{1 + A},\\
\Gamma_{31}^2&= -\frac {1}{2} \,\frac {A + A\,k
 + \cos^2\beta\,k - {\cos^2}\beta + 2}{1 + A}, \\
\Gamma_{32}^1&=\frac {1}{2} \frac{(1+k)(-A\sin^2\theta-1)+
(1-k)(cos^2\beta\cos^2\theta+\cos^2\theta-\cos^2\beta)}{1+A},\\
\Gamma_{32}^2&=\frac {1}{2}\frac{-2(1+k)(1+A)\sin(\psi-\alpha)\cos\theta+
(1-k)\sin\theta\sin\beta\cos\theta}{1+A},\\
\Gamma_{33}^3&= -(1+k) {\sin}(\psi-\alpha)\,{\cos}\theta
\endaligned
$$
Other $\Gamma $ are equal to zero.
Some components of the Shouten tensor different from zero are:
$$
\aligned
\overset 0 \to K_{121}^1&=-\overset 0 \to K_{122}^2=
 \, \frac {((k-1)^2 A+ 4\,k^{2})\,
\sin\beta\,\cos\beta\,\sin\theta}{4(1+A)^2}\\
\overset 0 \to K_{121}^2&=-\frac{ ( 1+k^2)( A^{2} +
A\,\cos^2\beta) +4Ak(1+k)+2k(A^2-A\,\cos^2\beta + 2\,k\,\cos^2\beta)}
{(1 + A)^2}\\
\overset 0 \to K_{132}^2&=\overset 0 \to K_{231}^3=
  \frac {( - 5\,A + 2
\,A\,k + 3\,A\,k^{2} - 4)\,\cos\beta\,\sin\beta\,\sin{\theta}}{4(1 +A)^2}\\
\overset 0 \to K_{133}^2&=
- \frac {( - 1 + k^{2})\,\sin\theta\,\sin\beta\,\cos\beta}{1 + A}
\endaligned
$$
The following components of the Shouten tensor are zero:
$$
\aligned
\overset 0 \to K_{121}^3&=\overset 0 \to K_{122}^3=
\overset 0 \to K_{123}^1=\overset 0 \to K_{123}^2=
\overset 0 \to K_{123}^3=\overset 0 \to K_{131}^1=
\overset 0 \to K_{131}^2=\overset 0 \to K_{132}^1=\\
\overset 0 \to K_{132}^2&=\overset 0 \to K_{133}^3=
\overset 0 \to K_{231}^1=\overset 0 \to K_{231}^2=
\overset 0 \to K_{232}^1=\overset 0 \to K_{232}^2=
\overset 0 \to K_{233}^3=0
\endaligned
$$

Expansion of the metric is given by the following formulae:
$$
\aligned
g^{44}&=\frac{2k^2}{A(A+1)^3\cos^2\beta\cos^2(\psi-\alpha)},\\
g^{45}&=\frac{-2k^2\sin\beta\sin(\psi-\alpha)}{A(A+1)^3\sin\theta\cos\beta
\cos(\psi-\alpha)},\\
g^{55}&=\frac{k^2(1-\cos^2\beta\sin^2(\psi-\alpha))}{A(1+A)^3\sin^2\theta
cos^2(\psi-\alpha)}.
\endaligned
$$

One of the components of the Wagner tensor is:
$$
K_{133}^1=
\frac{\sin^2\theta\cos^2\beta(k^2(A+4\sin^2\beta)+2Ak+A+4\cos^2\beta)}{4(1+A)}
$$
\medskip

From the last formula we get
\medskip

{\bf Theorem 1.} {\it  For any $k$ the Wagner curvature tensor is different from zero.}
\medskip

\

\centerline {\bf Conclusion}

\

From the Theorem 5.1, it follows that the Wagner tensor is
essentially different from the tensors constructed by Cartan [Ca]
and Agrachev's school [AS, AZ1, AZ2], since it doesn't recognize
the nilpotent case. A natural question is to find the
theory of Jacobi fields which corresponds to the Wagner curvature.

At the end let us note that the paper [Ta] appeared very recently, dealing
with geometrization of nonholonomic mechanics, based on some later Cartan's
work. The connections studied in [Ta] are generally not torsion-less.

\

\

{\bf Acknowledgement }  The research of both authors was partially
supported by the Serbian Ministry of Science and Technology
Project No  1643. One of the authors (V. D.) has a  pleasure to
thank Professor A. Agrachev and Dr I. Zelenko  for helpful
observations; his research was partially supported by SISSA
(Trieste, Italy)

\.

\

\centerline {\bf REFERENCES}

\

\item {[AS]} A. A. Agrachev, Yu. L. Sachkov, {\it Lectures on geometric control theory},  Trieste, SISSA preprint, 2001.

\item {[AZ1]} A. Agrachev, I. Zelenko, {\it Geometry of Jacobi Curves. I}, J. Dynamical and Control Systems, {\bf 8}, (2002), p. 93-140.

\item {[AZ2]} A. Agrachev, I. Zelenko, {\it Geometry of Jacobi Curves. II},
 SISSA preprint 18/2002/M

\item {[Ca]}{\rm E. Cartan, {\it Les syst\` emes de Pfaff \` a cinq variables et les \' equations aux d\' eriv\' ees partielles du second ordre}, Ann. Ec. Normale, {\bf 27}, (1910), p. 109-192.}

\item{[Hr]} {\rm H. Hertz, {\it Die Principien der Mechanik}, Leipzig, 1894.}

\item {[Ju]} {\rm V. Jurdjevic, {\it Geometric control theory}, Cambridge University Press, 1997.}

\item {[Mn]} {\rm R. Montgomery, {\it A Tour of Sub - Riemannian Geometries, their Geodesics and Applications}, Mathematical Surveys and Monographs, {\bf 91}, AMS, Providence, 2002.}

\item {{\rm [Sh]}} J. A. Shouten, {\it On nonholonomic connections}, Koninklijke akademie van wetenshappen te Amsterdam, Proceeding of sciences, (1928), {\bf 31}, No. 3, p. 291-298.

\item {[{\rm Sy}]} {\rm J.  L. Synge, {\it Geodesics in nonholonomic geometry}, Math. Ann. (1928), p. 738- 751.}

\item {[Ta]} {\rm J. N. Tavares, {\it About Cartan geometrization of non-holonomic mechanics}, Journal of Geometry and Physics, {\bf 45}, (2003), p. 1-23.}

\item {{\rm [Vr]}} G. Vranceanu, {\it Parallelisme et courlure dans une variete non holonome.} Atti del congresso Inter. del Mat. di Bologna, (1928).

\item {[AKN}]} {\nf Arnol\m d V.I., Kozlov V.V., Neishtad : {\rit Matematicheskie
aspekty klassichesko{\ji} i nebesno{\ji} mehaniki}, Dinamicheskie sistemy
{\it III}, Sovr. prob. mat., Fundamental\m nye napravleni\ja, Viniti, Moskva,
1985.}

\item {[{\rm Wa1}]} {\nf Vagner V.V.: {\rit Differencial\m a{\ja} geometri{\ja}
negolonomnyh mnogoobrazi{\ji}}, Tr. se\-minara po vekt. i tenzor. analizu,
v. {\rm II-III}, 1935., 269-315.}

\item {[{\rm Wa2}]} {\nf Vagner V.V.: {\rit Geometricheska{\ja} interpretaci{\ja} dvizheni{\ja}
negolonomnyh dina\-micheskih sistem}, Tr. seminara po vekt. i tenzor. analizu,
v. {\rm V}, 1941., 301-327.}

\item {[{\rm Wa3}]} {\nf Vagner V. V. {\rit Differencial\m a{\ja} geometri{\ja}
negolonomnyh mnogoobrazi{\ji}} $VIII$ mezhd. konkurs na soiskanie premii im. N. I. Lobachevskogo, (1937).

\item {[{\rm VG}]} {\nf Vershik A. M., Gershkovich V.{\JA}.: {\rit Negolonomnye dinamicheskie
sistemy. Geometrija raspredeleni{\ji} i variacionnye zadachi, Dinamicheskie
sistemy {\it VII}}, Sovr. prob. mat., Fundamental\m nye napravleni\ja,
Viniti, Moskva, 1987.}

\item {[{\rm Go}]} {\nf Gorbatenko E.M.: {\rit Differencial\m
na{\ja} geometri{\ja} negolonomnyh mnogoobrazi{\ji} (po V.V.
Vagneru)}, Geom. sb. Tomsk. u-ta, 1985., 31-43.}

\item {{\rm [NF]}} {\nf Ne{\ji}mark {\JU}.I., Fufaev N.A.: {\rit Dinamika
negolonomnyh sistem}, Nauka, Moskva, 1967.}

\end